\documentclass[aps, prb, floatfix, twocolumn, a4paper]{revtex4-1}

\usepackage{amsmath,   amsfonts, amssymb, graphicx,  graphics} 
\usepackage{epstopdf,  csquotes, color,   textcomp,  bm}
\usepackage{subfigure, array,    xspace,  bold-extra} 
\usepackage[squaren,   Gray,     cdot]{SIunits}

\usepackage[linktocpage=true, colorlinks=true,   urlcolor=blue,   linkcolor=blue,
            citecolor=blue,   pdfstartview=FitV, bookmarks=false, pdftoolbar=false]{hyperref}  
\newcommand{\fig}[1]{Fig.~\ref{fig:#1}}

\newcommand{\etal}{\emph{et al.~}}

\graphicspath{{../figures/}}

 \makeatletter \let\ps@titlepage\ps@plain \makeatother

\usepackage{fancyhdr}  

\begin{document}

\title{In-plane backward and Zero-Group-Velocity guided modes in rigid and soft strips}
\author{J\'er\^ome Laurent}  
\author{Daniel Royer}        
\author{Claire Prada}        \email{claire.prada@espci.fr}
\affiliation{ESPCI Paris, PSL Research University, CNRS, Institut Langevin, UMR 7587, 1 rue Jussieu, F-75005 Paris, France}
\date{\today}

\begin{abstract}
Elastic waves guided along bars of rectangular cross section exhibit complex dispersion. This paper studies in-plane modes propagating at low frequencies in thin isotropic rectangular waveguides through experiments and numerical simulations. These modes result from the coupling at the edge between the first order shear horizontal mode $SH_0$ of phase velocity equal to the shear velocity $V_T$ and the first order symmetrical Lamb mode $S_0$ of phase velocity equal to the plate velocity  $V_P$. In the low frequency domain, the dispersion curves of these modes are close to those of Lamb modes propagating in plates of bulk wave velocities $V_P$ and $V_T$. The dispersion curves of backward modes and the associated ZGV resonances are measured in a metal tape using non-contact laser ultrasonic techniques. Numerical calculations of in-plane modes in a soft ribbon of Poisson's ratio $\nu \approx 0.5$ confirm that, due to very low shear velocity, backward waves and zero group velocity modes exist at frequencies that are hundreds of times lower than ZGV resonances in metal tapes of the same geometry. The results are compared to theoretical dispersion curves calculated using the method provided in Krushynska and Meleshko (\href{https://doi.org/10.1121/1.3531800}{J.~Acoust.~Soc.~Am {\bf 129}, 2011)}.

\end{abstract}


\maketitle

\section{Introduction}

Early studies on the propagation of elastic waves in thin elongated strips were motivated by the need for long time ultrasonic delay lines. Today, rectangular elastic waveguides are encountered in miscellaneous contexts at different scales, from civil engineering structures to cantilever, micro-electromechanical systems (MEMS), or even in nanotechnology.\citep{Santamore02,Jean15} Thin elastic waves guides are also present in biological systems as for example, tendons and ligaments in the musculosquelettal system~\cite{Brum14} or the basilar and the tectorial membranes in the cochlea.\cite{sellon2015} Elastic waves propagation in these elongated structures is quite complex, several modes can propagate with interesting dispersion properties that strongly depend on the material Poisson's ratio. In particular, some modes exhibit negative dispersion.

The existence of backward elastic modes propagating in homogeneous waveguides is well known and has been put in evidence in the past century.\citep{Mindlin55,*Mindlin06} Experimental observations of these modes were first reported in wires~\citep{Meeker64,Meitzler65} and then for plates.\cite{Wolf88} At the minimum frequency of a backward mode, the resonance associated to the zero group velocity (ZGV) mode was pointed out by Tolstoy and Usdin~\cite{TolstoyUsdin57} and observed by Holland and Chimenti using air coupled transducers.\cite{Holland03} Then non-contact laser ultrasonic techniques allowed precise measurements and thorough analysis of these modes in metal plates,\citep{Prada05a,Balogun07,Clorennec06,Grunsteidl16} but also in circular hollow cylinders~\citep{Clorennec07,Ces11} and rods.\cite{Laurent14}\\

While elastic guided modes in isotropic plates and circular cylinders are well described by an analytical formulation of the dispersion equations, the theory of elastic rectangular waveguides is much more complex. Indeed, the modes cannot be expressed in a closed form, except for some particular width-to-thickness ratio that were determined by Mindlin and Fox.\cite{Mindlin60} Since the seminal work of Morse,\cite{Morse50} several low frequencies approximations were proposed, for example, Medick used a one dimensional quadratic approximation allowing the calculation of the first eight breathing modes.\cite{Medick68} More recently, several studies based on numerical simulations using finite and boundary element methods provided approximate dispersion curves.\citep{Mukdadi02,Mukdadi03,Stephen04,Cortes10,Cegla08,Kavzys16} A complete review of the theory can be found in the papers by Meleshko and Krushynska~\citep{Meleshko10,Krushynska11a} where the dispersion equations of rectangular elastic bars are formulated as infinite series and normal modes are studied for various width to thickness ratio and material parameters. The existence of several backward modes for Poisson's ratio up to $0.5$ was pointed out for bars of width-to-thickness ratio higher than $5$ in the study by Krushynska \etal (section III of paper~\cite{Krushynska11a}).

In addition to these theoretical studies, the large majority of experimental results found in the literature deals with flexural modes at low frequencies. Seventy years ago, Morse measured dispersion curve on bars of different width-to-thickness ratios and compared them with low frequency approximation.\cite{Morse48} Twenty years later, the dispersion curves of higher order modes including backward modes were measured by Hertelendy in a square bar of an aluminum alloy.\cite{Hertelendy68} More recently, several dispersion curves were measured in wooden bars by Veres \etal using laser interferometry and then compared to numerical simulations of a wooden orthotropic beam.\cite{Veres04} Lately, Serey~\etal addressed the selective generation of pure guided waves in an thin metal bar of rectangular cross-section using three component interferometry and an array of transducers, in a frequency range below the shear cut-off frequency.\cite{Serey18} They were able to detect and generate modes with dominant out of plane displacement corresponding to edge reflections of the $A_0$ Lamb mode. In picosecond acoustics, Jean~\etal have measured elastic resonances and a backward mode in gold nano-beam of rectangular cross section ($370$-nm wide and $110$-nm thick). With semi-analytical finite element simulations (SAFE), the backward wave was identified as the first dilatational mode.\cite{Jean15} Very few studies report on measurments of in-planes modes. Cegla~\etal detected shear horizontal mode propagating in a steel tape with a dual laser probe at frequency far above the shear cut-off frequency where the mode is non-dispersive and propagates at the shear wave velocity.\cite{Cegla08} In-plane guided waves were also observed by Sellon~\etal in ex-vivo tectorial membranes using optical means.\cite{sellon2015}\\

In the present paper, we consider the low frequency in-plane modes propagating in thin rectangular bars. We study the corresponding backward mode and associated ZGV resonances for metal tapes as well as soft ribbons. The low frequency in-plane guided modes of rectangular tapes are described in section~\ref{sec1}. The dispersion curves are calculated for a metal and a soft strip following Krushynska~\etal\cite{Krushynska11a} and compared to the corresponding Rayleigh-Lamb (RL) modes approximation. In section~\ref{sec2}, ZGV resonances and dispersion curves of in-plane modes measured using laser ultrasonic techniques are presented and compared to the theory. In the last section~\ref{sec3}, comparison between modes propagating in hard and soft ribbons is achieved through numerical simulations. 

\section{In-plane modes of a thin rectangular beam} \label{sec1}

At low frequencies, an isotropic plate supports the propagation of only three modes : the shear horizontal mode $SH_0$, the compressional symmetric $S_0$ and the flexural antisymmetric $A_0$ Lamb modes. The symmetrical $SH_0$ and $S_0$ modes are both in-plane and non-dispersive: while the phase velocity of $SH_0$ is the shear velocity $V_T$, the $S_0$ phase velocity is approximated by the plate velocity $V_P$ expressed from the bulk velocities $V_L$ and $V_T$, as 
\begin{equation}
V_P = 2V_T\sqrt{1-\frac{V_T^2}{V_L^2}}.
\end{equation}
This approximation is valid for frequencies $f$ smaller than the first shear cut-off frequency $V_T/(2d)$, where $d$ is the plate thickness.\cite{Royer99a} 

If the plate is bounded in the transverse dimension, thus forming a rectangular waveguide of width $b \gg d$, then the in-plane modes $SH_0$ and $S_0$ are coupled through reflections at the edges (Fig.~\ref{fig:InPlaneModes}). As shown by Morse~\cite{Morse50} and by Medick and Pao,\cite{Medick65} this coupling is very similar to the one of shear and compressional bulk waves in a plate. The resulting guided modes have dispersion curves close to those of Lamb modes in a plate of thickness $b$ and of bulk wave velocities $V_P$ and $V_T$ which corresponds to a material of Poisson's ratio 
\begin{equation}
\nu' = \frac{\nu}{1+\nu}, 
\label{eq_nup}
\end{equation}
where $\nu$ is the Poisson's ratio of the beam material.

\begin{figure}[!ht]
\centering
\includegraphics[width=\columnwidth]{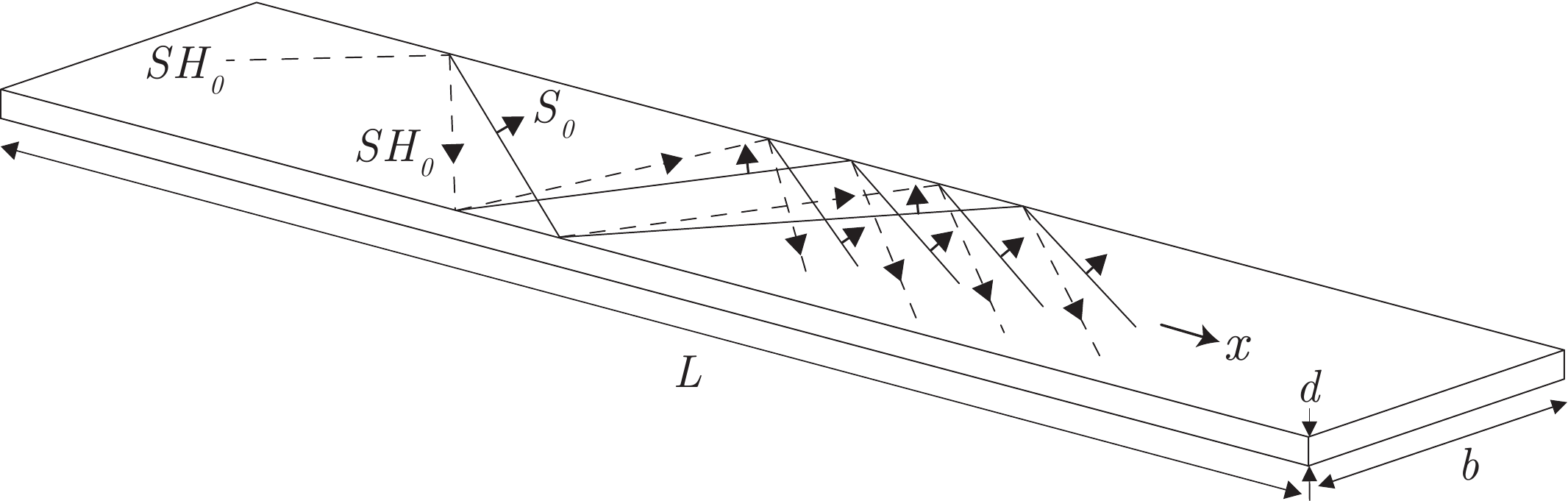}\label{fig:InPlaneModes} 
\caption{In-plane modes conversion in a thin rectangular beam.}
\label{fig:InPlaneModes}
\end{figure}

The calculation of guided modes in a rectangular waveguide~\cite{Krushynska11a} is described in Krushynska \etal Like in the earlier paper by Fraser,\cite{Fraser69} the modes are classified in four families depending on the displacement symmetries: the longitudinal $L$-modes, the torsional $T$-modes, the bending $B_x$-modes and $B_y$-modes. For thin beams with the smallest dimension along $y$-axis, the symmetrical in-plane $S'$-modes correspond to the longitudinal $L$-modes and the anti-symmetrical in-plane $A'$-modes to the bending $B_x$-modes. In Krushynska \etal,\cite{Krushynska11a} the low frequency approximation of in-plane modes was confirmed for a plate of width to thickness ratio $b/d=5$ and of Poisson's ratio $\nu=1/4$, and for frequencies up to four times the first cut-off frequency $V_T/(2b)$.

The first symmetrical mode, here denoted $S'_0$, is non dispersive below the first cut-off frequency and has a phase velocity $V'_P$ slightly lower than the plate velocity $V_P$ and equal to 
\begin{equation}
V'_P = 2V_T\sqrt{1-\frac{V_T^2}{V_P^2}}.
\end{equation}
Using the plate to shear velocity ratio 
\begin{equation}
\frac{V_P}{V_T} =\sqrt{\frac{2}{1-\nu}}= \sqrt{\frac{2(1-\nu')}{1-2\nu'}}, 
\label{eq_VP1}
\end{equation}
a remarkably simple formulation of the pseudo-plate velocity to shear velocity ratio is obtained
\begin{equation}
\frac{V'_P}{V_T} = \sqrt{\frac{2}{1-\nu'}}=\sqrt{2(1+\nu)}.
\label{eq_VP2}
\end{equation}
In fact, the velocity $V_P'$ is the bar velocity $\sqrt{E/\rho}$, which is the velocity of the low frequency longitudinal mode in an elastic guide of finite cross-section as established by Lord Rayleigh.\cite{Kynch57} Noteworthy simple expressions of the velocities are found for two particular cases. For $\nu=1/3$ the plate velocity is $V_P=\sqrt{3}V_T$ then $\nu'=1/4$ and the velocity of the $S'_0$ mode at low frequencies is $V'_P=\sqrt{8/9}V_P \approx 0.94V_P$. For soft materials $\nu$ approaches $0.5$, the plate velocity is $V_P = 2 V_T$ which is much smaller that the longitudinal velocity.\cite{Sarvazyan75} Thus, in soft ribbon the bar velocity is $V'_P=\sqrt{3} V_T$. These remarkable parameters are gathered in Table \ref{vitesse_de_plaque}. 
\begin{table}[!ht]
\begin{minipage}[c]{0.8\columnwidth}
\begin{ruledtabular}
\centering
\caption{\label{vitesse_de_plaque} Remarkable plate and bar velocities.}
\begin{tabular}{lcccc}
\textbf{Material} & ~$\nu$~          & ~$\nu'$~         & ~$V_P$~          & ~$V'_P$~                   \\
\hline
\textbf{Metal}    & $\frac{1}{3}$    & $\frac{1}{4}$    & $\sqrt{3}V_T$    & $\sqrt{\frac{8}{3}}V_T$    \\
\textbf{Soft}     & $\frac{1}{2}$    & $\frac{1}{3}$    & $2V_T$           & $\sqrt{3}V_T$              \\
\end{tabular}
\label{table:Resume}
\end{ruledtabular}
\end{minipage}
\end{table}
\begin{figure*}[ht]
\subfigure{\includegraphics[width=0.9\textwidth]{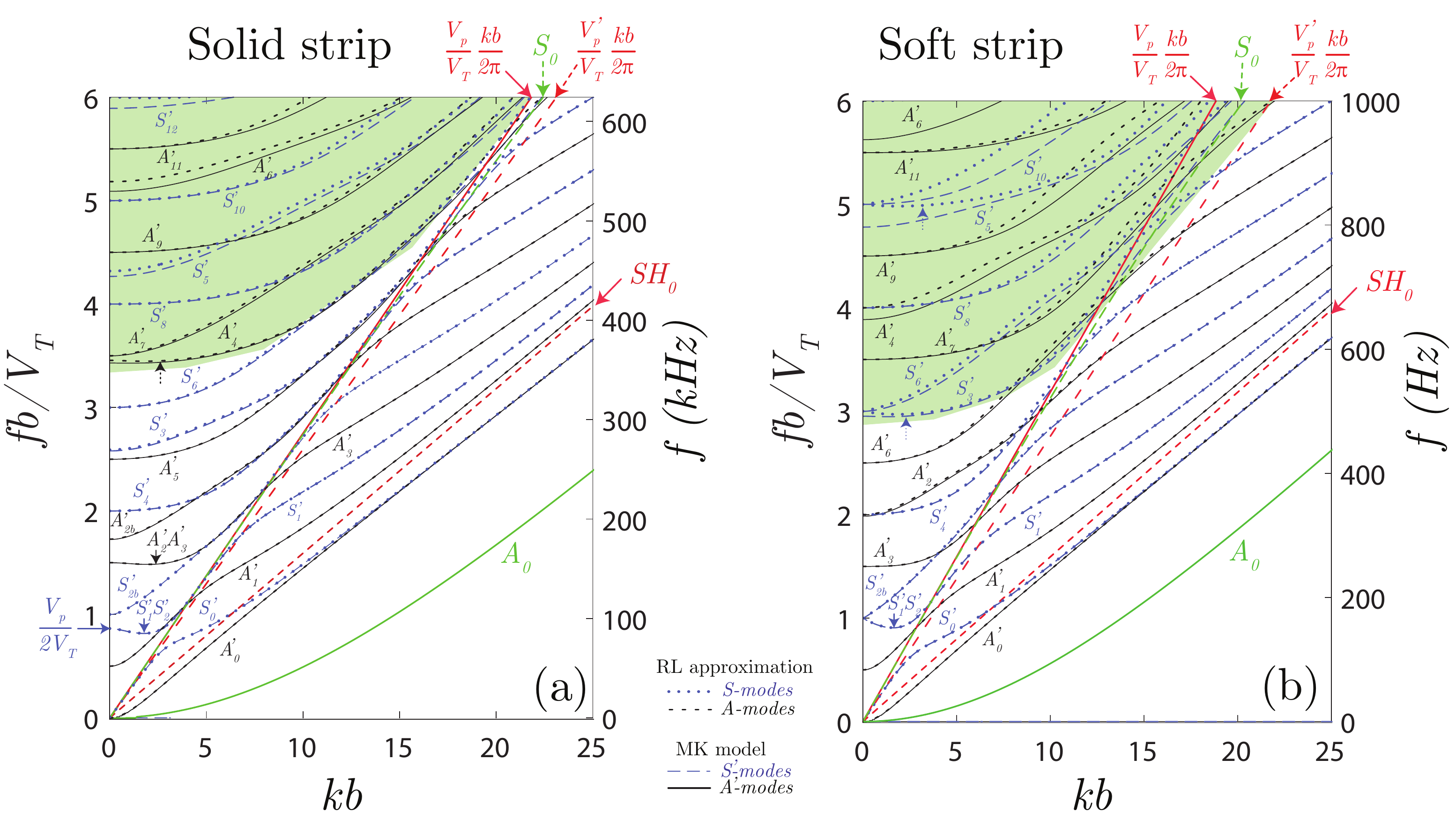}\label{fig:Th_Disp_CurvesSofta}} 
\subfigure{\label{fig:Th_Disp_CurvesSoftb}}
\caption{In-plane modes dispersion curves calculated with MK-model for solid and soft strips of width $b=30$~mm, thickness $d=2$~mm and Poisson's ratio $\nu$, compared to the Rayleigh-Lamb modes for a plate of thickness $b$ and Poisson's ratio $\nu'=\nu/(1+\nu)$. Thick red, blue and green solid lines are the first two Lamb modes and the first shear mode of a plate of thickness $d$ and Poisson's ratio $\nu$; the red dashed line indicates the plate velocity. (a) Duraluminum tape of Poisson's ratio $\nu=0.34$, (b) soft ribbon of bulk wave velocities $V_L=1500$~m/s and $V_T=5$~m/s. The cut-off frequency of $S'_1$ mode is $fb=V_T$. The $S_0$ mode coincides with the straight line $2\pi f=V_P k$ while the $S'_0$ mode is close to the line $2\pi f = V'_P k$ for $fb<0.4 V_T$.}
\end{figure*}

Here, using the Meleshko Krushynska (MK) model,\cite{Krushynska11a} theoretical dispersion curves are calculated for a Duraluminum tape of thickness $d=2$~mm, width $b=30$~mm and bulk velocities $V_L=6400$~m/s, $V_T=3140$~m/s. In Figure~\ref{fig:Th_Disp_CurvesSofta}, the normalized frequency $fb/V_T$ is represented as a function of the dimensionless wave number $kb$. Lamb modes in a plate of bulk velocities $V_T$ and $V_P$ and thickness $b$ are also displayed for comparison. A very good agreement with the Lamb modes approximation is observed for frequencies up to $f=4V_T/b$. As in a plate, several backward and ZGV modes exist. They occur when a shear (or $SH_0$) cut-off frequency $nV_T/b$ is close to a plate mode (or $S_0$) cut-off frequency $mV_P/b$ of the same symmetry. $S'_1S'_2$- and $A'_2A'_3$-ZGV points are found at frequency wave number $(fb/V_T,kb) = (0.824, 1.77)$ and $(1.49, 1.95)$ respectively [Figure~\ref{fig:Th_Disp_CurvesSofta}]. The thick solid lines represent the first modes $A_0$, $S_0$, and $SH_0$ of an unbounded plate of thickness $d$. For higher frequencies, between $A_0$ and $SH_0$ modes, the $A'_0$ and $S'_0$ modes converge towards the edge or pseudo-Rayleigh mode.\\

While in plates, the $S_1S_2$-ZGV Lamb mode only exits for Poisson's ratio below $0.451$,\cite{Prada08a} in thin rectangular beams, the $S'_1S'_2$-ZGV mode exists for any Poisson's ratio. The behaviour of in-plane modes is particularly interesting in the case of soft material.
Indeed, for $\nu \approx 0.5$, the plate velocity $V_P$ is equal to $2V_T$ which results in a Dirac cone at the frequency $f=V_T/b$ as it happens for Lamb modes in a plate of Poisson's ratio $\nu=1/3$.\citep{Mindlin55,*Mindlin06,Maznev14,Stobbe17} This is illustrated in Figure~\ref{fig:Th_Disp_CurvesSoftb} displaying the in-plane mode dispersion curves for a $2$~mm thick and  $20$~mm wide tape of velocities $V_L=1500$~m/s and $V_T=5$~m/s ($\nu= 0.4998$). Such low shear velocity corresponds to typical velocities measured in agar gel or soft tissues. The second $SH_0$ cut-off frequency coincides with the first $S_0$ cut-off frequency $f=V_P/(2b)$ and the repulsion between the branches $S'_1$ and $S'_2$ results in a $8\%$ relative bandwidth $S'_{2b}$ backward branch (from $230$ to $249$~Hz in the presented case). For both branches, the group velocity remains finite as $k \rightarrow 0$ and converges to $V_g = 2V_T/\pi$.\cite{Mindlin06} The $S'_1S'_2$-ZGV mode occurs at a frequency~$\times$~width product $fb=4.53$~kHz.mm which is more than two orders of magnitude lower than the frequency~$\times$~width product of the same ZGV mode in a metal tape $fb=2.64$~MHz.mm.

\begin{figure}[!ht]
\centering
\subfigure{\includegraphics[width=\columnwidth]{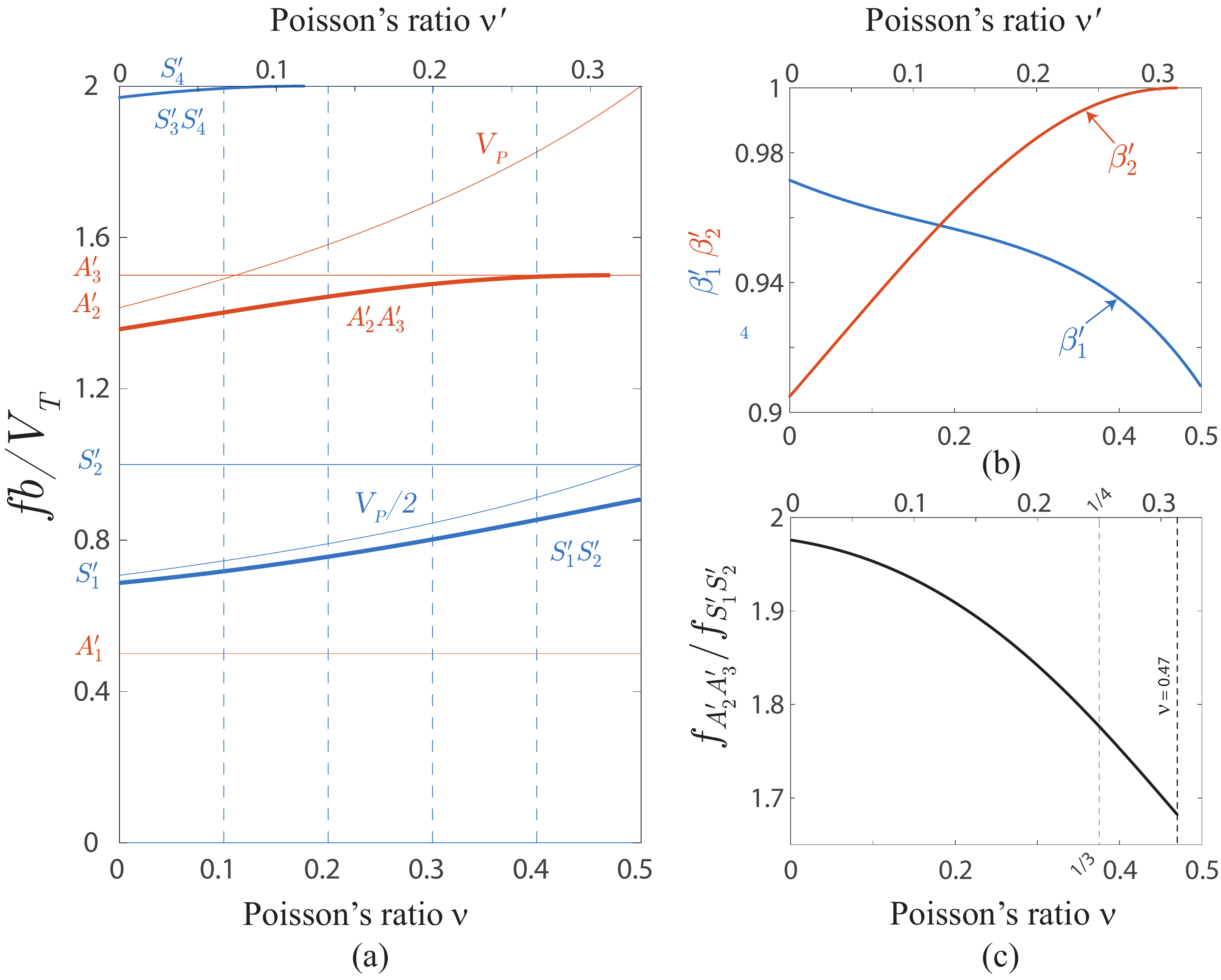}\label{fig:CutoffZGVbetaa}}
\subfigure{\label{fig:CutoffZGVbetab}} 
\subfigure{\label{fig:CutoffZGVbetac}}
\subfigure{\label{fig:CutoffZGVbetad}} 
\caption{In-plane modes of a thin ribbon: (a) Normalized cut-off (thin lines) and ZGV frequencies (thick lines) as a function of the material Poisson's ratio; the thin horizontal lines correspond to $SH_0$ cut-off frequencies, (b) Variations of the resonance parameters $\beta'_1$ and $\beta'_2$ versus Poisson's ratio, (c) Ratio between $A'_2A'_3$- and $S'_1S'_2$-ZGV frequencies.}
\label{fig:CutoffZGVbeta}
\end{figure}

As for plates or cylinders, the existence of ZGV modes depends on the material elastic parameters. For isotropic thin rectangular beams, this dependence is illustrated in Figure~\ref{fig:CutoffZGVbetaa} displaying cut-off and ZGV resonance frequencies versus Poisson's ratio. Horizontal and increasing curved thin lines corresponds to $SH_0$ and $S_0$ width resonances, respectively. It appears that, while the $S'_1S'_2$-ZGV mode exists for all Poisson's ratio, the $A'_1A'_2$-ZGV mode exists for Poisson's ratio $\nu$ up to 0.47. Using the approach proposed in Clorennec \etal for plates,\cite{Clorennec07} the resonance parameter $\beta'_1$ is defined as the ratio of the $S'_1S'_2$-ZGV frequency to the first $S_0$ cut-off frequency $V_P/(2b)$ and $\beta'_2$ is defined as the ratio of $A'_1A'_2$-ZGV frequency to the third $SH_0$ cut-off frequency $3V_T/(2b)$. There is a one-to-one correspondence between ratio of the first two ZGV resonances $3\beta'_2V_T/(\beta'_1V_P)$ and the Poisson's ratio [Figure~\ref{fig:CutoffZGVbetaa}]. As a consequence and similarly to plates, the measurement of the first two local resonances provides the material Poisson's ratio of the rectangular tape.

\section{Measurements on a metal tape} \label{sec2}

In-plane guided modes are observed on metal strips using non-contact laser ultrasonic techniques. A Q-switched Nd:YAG (Yttrium Aluminium Garnet) laser providing $8$-ns pulses of $10$-mJ energy at a $8$-Hz repetition rate (Quantel Centurion) is used as a thermoelastic source. The laser beam of diameter $2.5$~mm is focused into a narrow line with a beam expander ($\times 4$) and a cylindrical lens (focal length $250$~mm). The full length of the source at $1/e$ of the maximum value was found to be $10$ mm and the width was estimated to be $0.3$~mm. With this features, the absorbed power density was below the ablation threshold. The source length is smaller than the optimal length to generate a ZGV mode but reasonable to generate several in-plane modes. \footnote{Optimal source length at constant maximum surface energy density: The absorbed energy distribution is written as $\bm{E}(x) = I\,\exp(-x^2/l^2)$ where $I$ is supposed to be lower than the ablation threshold. The resulting Fourier transform is given by $\bm{E}(k) = (I\sqrt{\pi}l) \times \exp(-l^2k^2/4)$. The amplitude of a mode of wave number $k$ reaches a maximum for the optimal source length $l_{opt} = \lambda/(\pi\sqrt{2})$.} 

\begin{figure*}[!ht]
\centering
\subfigure{\includegraphics[width=0.8\textwidth]{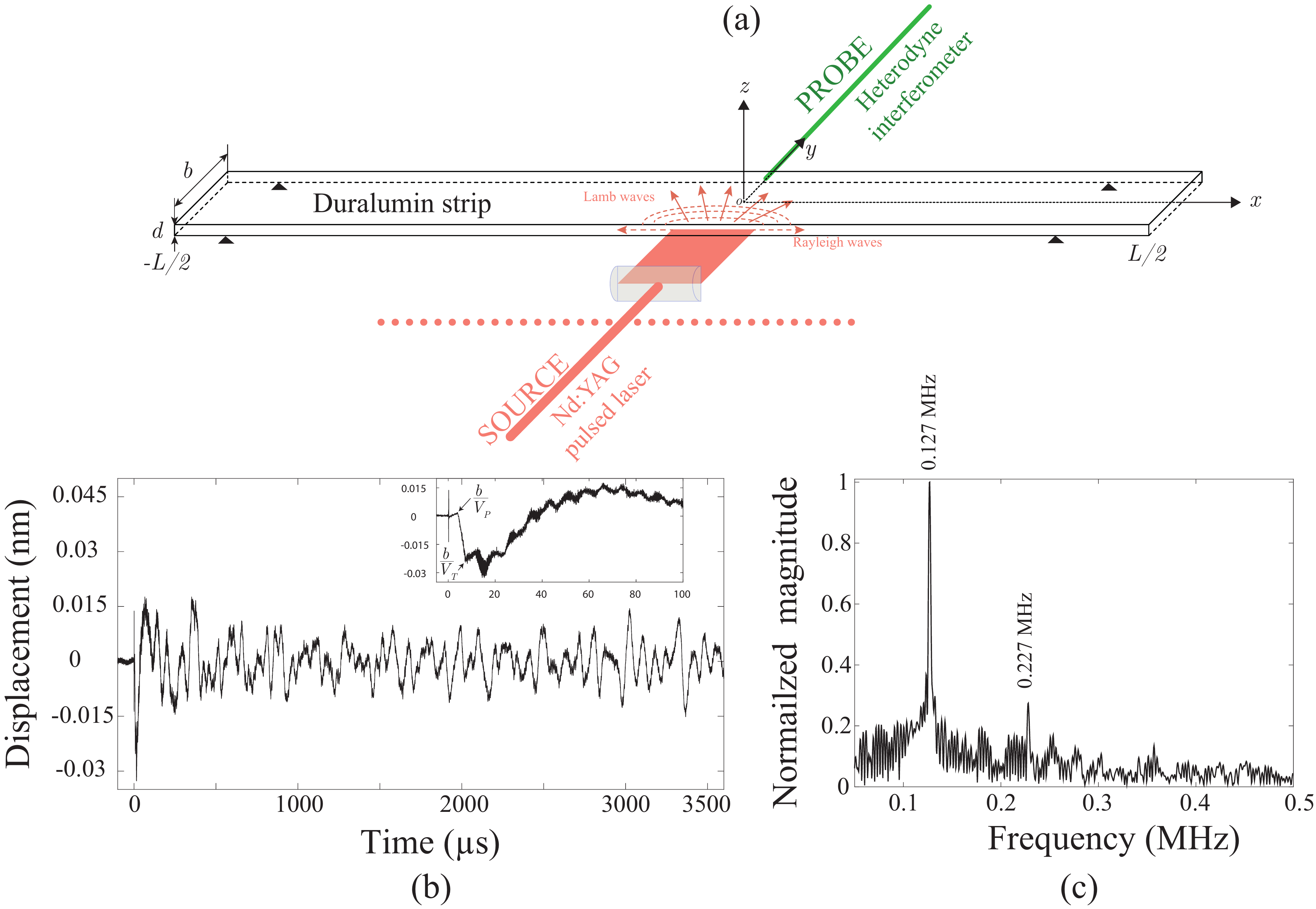}\label{fig:seta}}
\subfigure{\label{fig:setb}} 
\subfigure{\label{fig:setc}}
\caption{(a) Schematic diagram of in-plane guided wave measurement in a metallic strip. The excitation is achieved on one edge with a laser line source parallel to the plate surface. The normal surface displacement is measured on the opposite edge with an interferometer, (b) Displacement measured at the epicentre on a strip of thickness $d=0.5$~mm and width $b=20$~mm, (c) Fourier transform of the first $200~\micro$s of the measured signal.}
\label{fig:set}
\end{figure*}

In order to generate modes with symmetrical displacement component $u_z$, the line source was located in the middle of the tape edge (plane $z=0$).
Then, to detect in-plane modes, the normal surface displacement $u_y$ was measured on the opposite edge with a heterodyne interferometer equipped with a $100$-mW frequency doubled Nd:YAG laser (optical wavelength $\Lambda= 532$~nm) [\fig{seta}]. 

First measurements were done on a $1$-m long Duraluminum tape of thickness $d=0.5$~mm and width $b=20$~mm. A significant displacement is detected for more than $3$~ms after the laser impact~\fig{setb}. This duration is mostly due to the reflections at the tape ends of edge modes similar to those measured by Jia~\etal\cite{Jia96} During the first $100~\micro$s (inset) a typical waveform is observed with the first arrival of the compressional mode $S_0$, followed by the first arrival of the shear mode $SH_0$, as well as a low frequency resonance. The Fourier transform displayed in~\fig{setc} highlights the first two ZGV resonances: $S'_1S'_2$ at $0.127$~MHz and $A'_2A'_3$ at $0.227$~MHz.

\begin{figure}[!ht]
\centering
\subfigure{\includegraphics[width=\columnwidth]{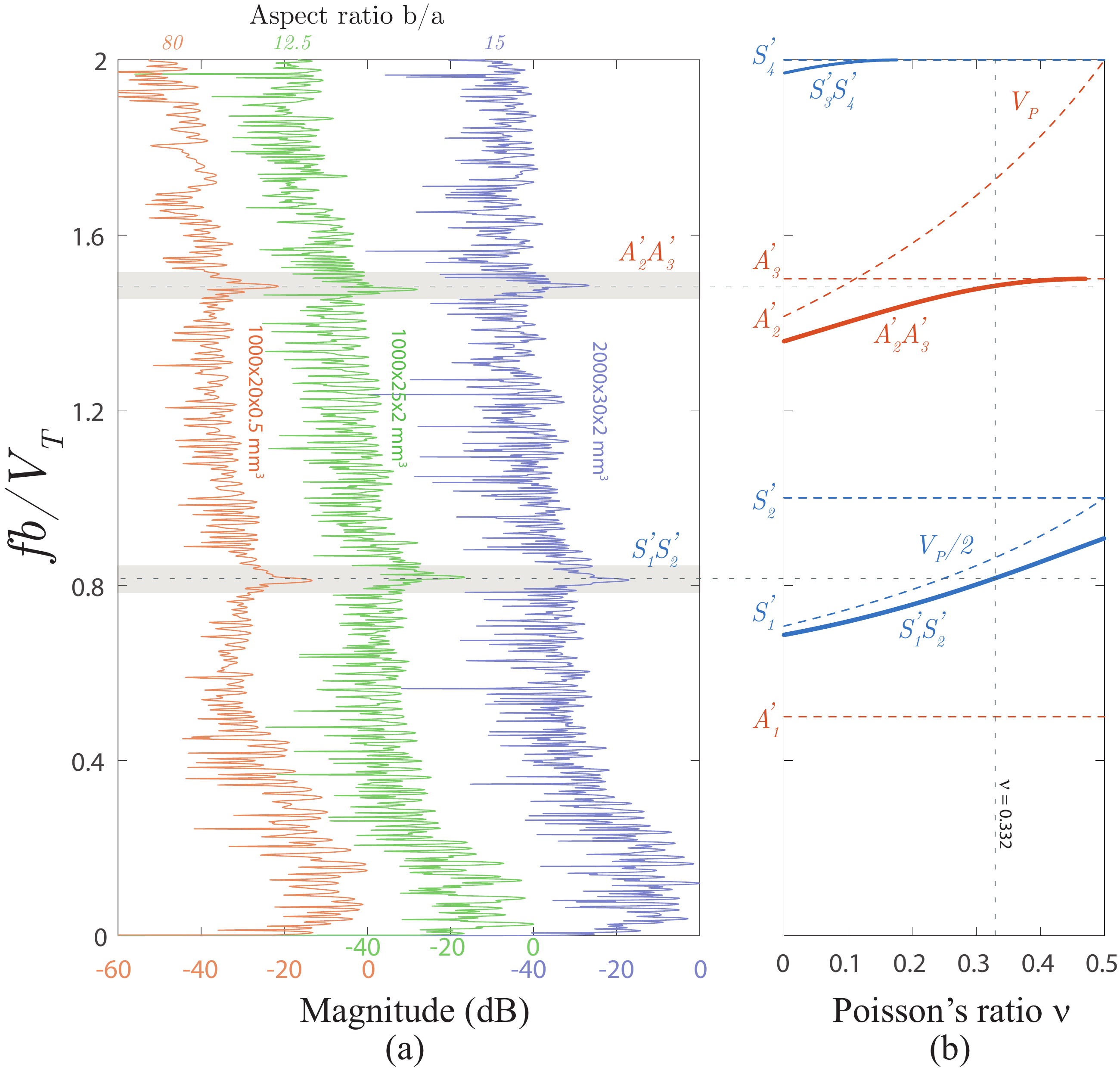}\label{fig:set4a}}
\subfigure{\label{fig:set4b}} 
\caption{(a) Local resonance spectrum measured on three Duraluminum tapes, (b) ZGV (solid lines) and cut-off (dashed lines) normalized resonance frequencies as a function of Poisson's ratio.}
\label{fig:set4}
\end{figure}

Similar measurements were also performed on two $d=2$~mm thick tapes of width $b=25$ or $30$~mm. Normalized spectrum are displayed in~\fig{set4a}. For the three tapes, the local resonance frequencies correspond to the theoretical $S'_1S'_2$- and $A'_2A'_3$-ZGV resonances frequencies calculated for a material of Poisson's ratio $\nu=0.332$ [\fig{set4b}] which is in good agreement with values generally found in the literature for aluminum alloys.

\begin{figure*}[!ht]
\centering
\subfigure{\includegraphics[width=0.8\textwidth]{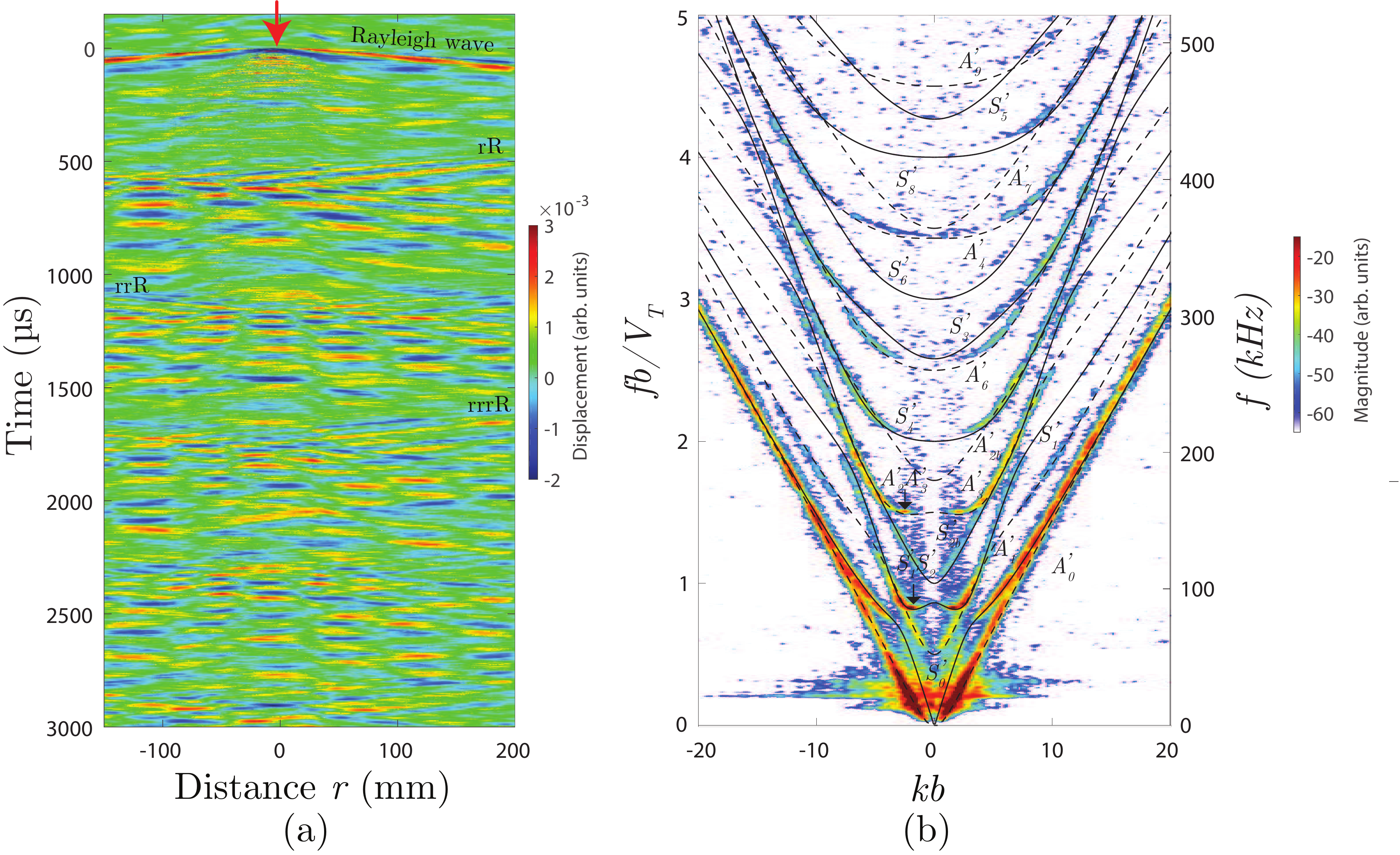}\label{fig:dispersionexpa}}
\subfigure{\label{fig:dispersionexpb}} 
\caption{(a) Displacement $u_y$ measured along the edge of a Duraluminum tape of dimensions $L=2$~m, $b=30$~mm and $d=2$~mm. (b) 2D-Fourier analysis of the displacement field with superimposed theoretical dispersion curves in black and red dashed lines.}
\label{fig:dispersionexp}
\end{figure*}

\begin{figure}[!ht]
\centering
\includegraphics[width=0.95\columnwidth]{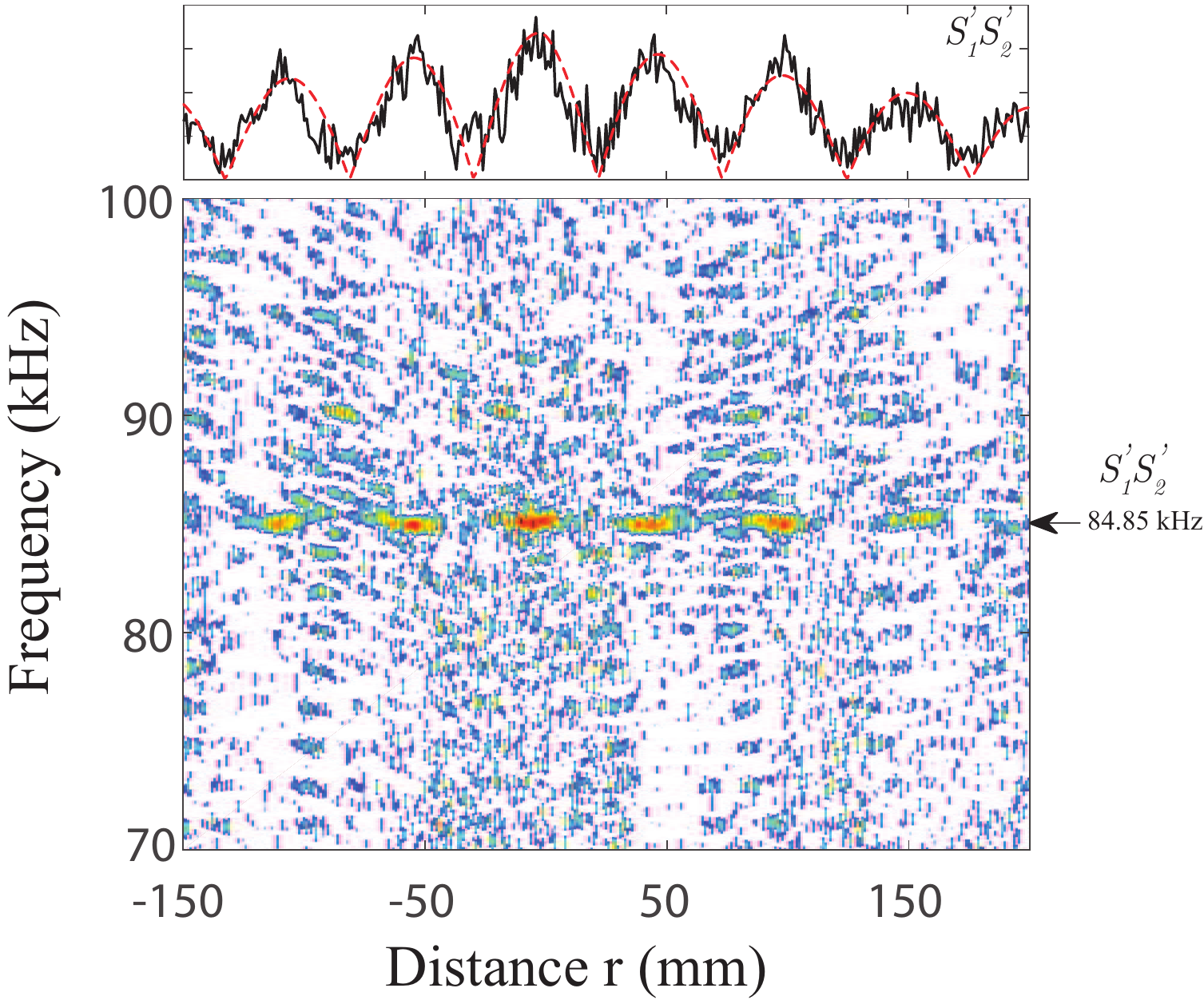}
\caption{Spatial distribution of the $u_y$ measured displacement around the $S'_1S'_2$-ZGV frequency. The top figure represents the amplitude profile at the frequency $84.85$~kHz (black) and the theoretical damped cosine profile (red dashed line). }
\label{fig:firstZGV}
\end{figure}

In order to confirm the origin of the resonances, the source was translated along the strip over $350$~mm to observe the different modes. A Bscan $u_y(x,t)$ was recorded during $3$~ms [\fig{dispersionexpa}]. The signal is dominated by the pseudo-Rayleigh waves (edge wave) and the first reflections at the tape end appear after $400~\micro$s. An average of 2D-Fourier transforms calculated on $450~\micro$s sliding time windows provides the dispersion curves. Several modes are measured up to a frequency~$\times$~width product of $fb=4V_T$. The $S'_1S'_2$- and $A'_2A'_3$-ZGV points just below the cut-off frequencies $fb=V_T$ and $fb= 1.5V_T$ as well as the associated backward modes are clearly identified and in good agreement with theoretical curves. The temporal Fourier transform of the displacement profile $u_y(x,f)$ is displayed around the $S'_1S'_2$-ZGV frequency (\fig{firstZGV}). The interference pattern of the ZGV modes is clearly observed at $f=84.85 kHz$ and shown in the top figure. The red dashed line is the damped cosine function $|\cos(2\pi x/\lambda_{S'_1S'_2})|\times\exp(-\alpha |x|)$ where the ZGV mode wavelength is $\lambda_{S'_1S'_2}=103$~mm and $\alpha$ is an adjusted damping parameter. This profile is in good agreement with the numerical displacements field that will be shown in the appendix.

\section{Hard and soft ribbon: comparison through numerical analysis} \label{sec3}

In order to complete these measurements, the finite-difference time domain code Simsonic~\cite{Bossy04} was used to simulate the propagation of low frequency in-plane modes in a thin tape. The first calculation was achieved for the Duraluminum tape similar to the one in the experiments ($L=10$~m, $b=30$~mm and $d=2$~mm). Then, we calculated the waves propagating in a soft ribbon of Poisson's ratio $\nu \approx 0.5$ with $V_T=5$~m/s, and $V_L=1500$~m/s and of geometrical parameters $L=2$~m, $b=20$~mm and $d=2$~mm. The strips lengths were chosen large enough to avoid mode reflections at the end.

\begin{figure*}[!ht]
\centering
\subfigure{\includegraphics[width=\textwidth]{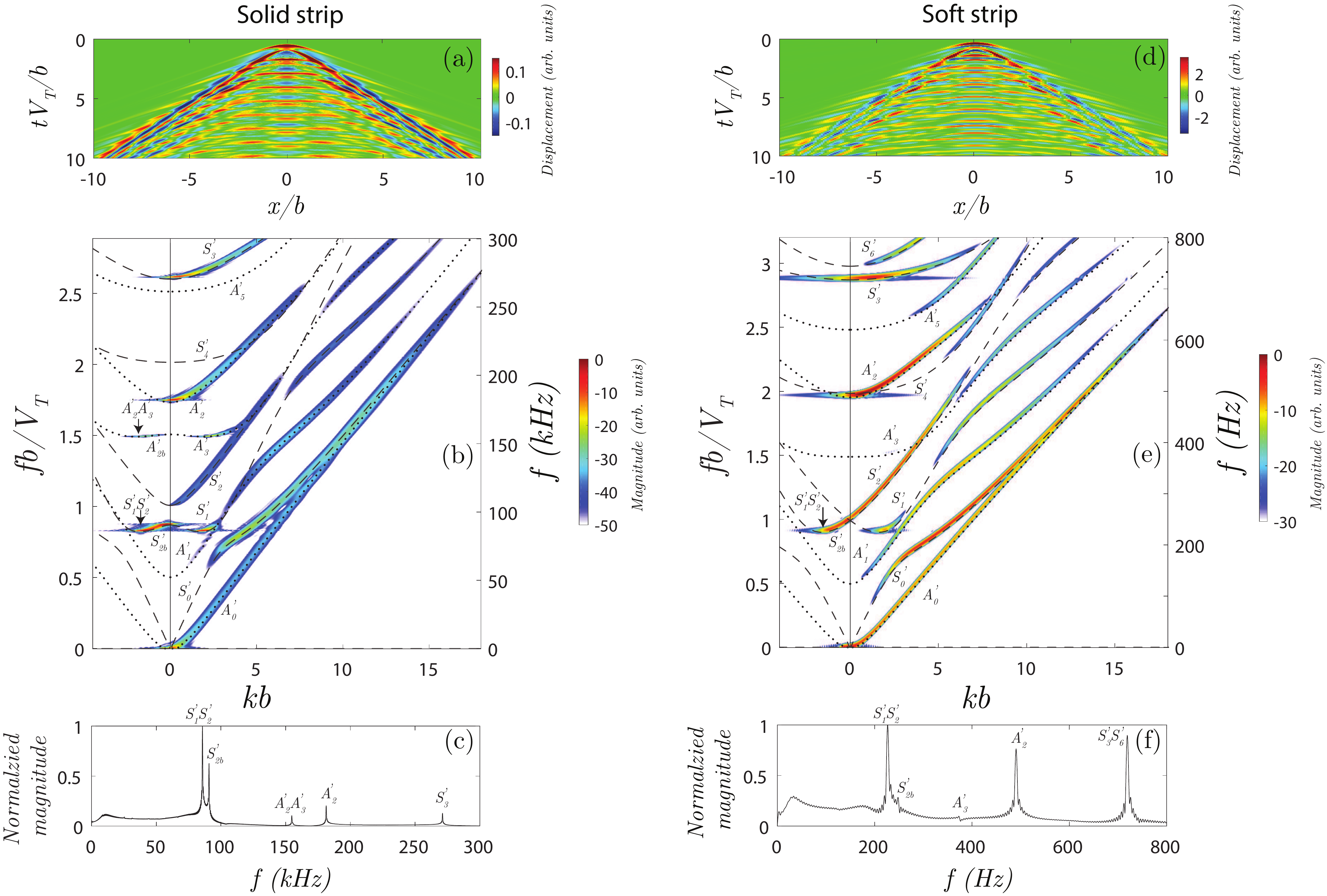}\label{fig:DispersionNumSolidSofta}}
\subfigure{\label{fig:DispersionNumSolidSoftb}} 
\subfigure{\label{fig:DispersionNumSolidSoftc}}
\subfigure{\label{fig:DispersionNumSolidSoftd}} 
\subfigure{\label{fig:DispersionNumSolidSofte}}
\subfigure{\label{fig:DispersionNumSolidSoftf}} 
\caption{Simulated in-plane mode propagation in a Duraluminum tape of parameters $L=10$~m, $b=30$~mm, $d=2$~mm, $V_L= 6400$~m/s, $V_T= 3140$~m/s (left) and a soft ribbon of parameters $L=2$~m, $b=20$~mm, $d=2$~mm, $V_L=1500$~m/s, $V_T=5$~m/s (right): normal edge velocity as a function of dimensionless time and distance (a) and (d); 2D-Fourier transform of the normal edge velocity with superimposed theoretical dispersion curves obtained with MK model (b) and (e); local resonances recorded at point $(x,y,z)=(0,-b/2,0)$ opposite to the source center (c) and (f).}
\label{fig:DispersionNumSolidSoft}
\end{figure*}

In order to excite in-plane modes, a normal velocity source with a constant profile along the thickness and a Gaussian profile along the $x$-axis, was applied on one tape edge. Then the velocity field $V_y(t,x,b/2,0)$ was registered along the opposite edge for $-1<x<1$~m, during $2$~ms for the metal tape and $200$~ms for the soft ribbon. It is represented as function of dimensionless time $tV_T/b$ and distance $x/b$ Fig.~\ref{fig:DispersionNumSolidSofta} and Fig.~\ref{fig:DispersionNumSolidSoftd}. In both Bscan, it can be observed that after two or three reflections of the $S_0$ mode, the stationary modes appear.

The 2D-Fourier transform of the field $V_y(t,x,b/2,0)$, $x>0$ is calculated and displayed with the theoretical dispersion curves calculated with the MK-model Fig.~\ref{fig:DispersionNumSolidSoftb} and Fig.~\ref{fig:DispersionNumSolidSofte}. The modes having a non-zero velocity component $V_y$ are observed and are in good agreement with the theoretical ones. For the solid strip, the two backward modes $S'_{2b}$ and $A'_{2b}$ are clearly detected. In the soft strip, the well generated $S'_{2b}$ backward mode appears as the continuation of the $S'_2$ mode, confirming the existence of a Dirac cone. 

The frequency spectra of the velocity recorded at point $(x,y,z)=(0,-b/2,0)$ opposite to the source center are plotted Fig.~\ref{fig:DispersionNumSolidSoftc} and Fig.~\ref{fig:DispersionNumSolidSoftf}. For the solid strip, the three resonances are associated to the $S_0$ mode cut-off of $S'_{2b}$, $A'_2$ and $S'_3$, the other two are the $S'_1S'_2$- and $A'_2A'_3$-ZGV resonances that were measured in the experiment. The cut-off frequency resonances are well excited in this numerical simulation because the source is normal to the edge, which is in favor the $S_0$ plate mode generation, while a laser source in the thermoelastic regime mostly generate a shear wave.\cite{hutchins81} For the soft strip, the $S'_1S'_2$- and $S'_3S'_6$-ZGV resonances and the $A'_2$ cut-off resonance dominate. On the contrary, there is no significant resonance at the $S'_2$ (or $S'_{2b}$) cut-off frequency. In fact, as shown by Mindlin,\cite{Mindlin06} at the Dirac cone the group velocity remains finite and is $V_g=2V_T/\pi$ so that all modes are propagative and no resonance occurs. 

\section{Conclusion}

The propagation of in-plane modes in thin rectangular beams, corresponding to low frequency longitudinal $L$-modes and bending $B_x$ modes, was studied. Dispersion curves were simulated and measured in metallic tape. A good agreement was observed with theoretical dispersion curves calculated with the Meleshko-Krushynska (MK) model. The backward modes and the associated Zero-Group-Velocity mode resonances were clearly observed. We showed that these resonances do not depend on the strip thickness and that the measurement of the two lowest ZGV resonances can be used to determine the strip material Poisson's ratio $\nu$. Non-dissipative soft ribbons were then considered, showing that backward modes exist at frequencies very low compared to the longitudinal cut-off frequency. These backward modes weakly coupled to the embedding fluid, could occur in the tectorial membrane for example and the associated in-plane displacements may play a role in the complex audition mechanism. However, theoretical studies should be conducted to account for viscoelastic properties and new experiments should be developed to measure guided waves in soft strips.\\

\section*{Acknowledgements} 

The authors wish to thank B. G\'erardin for drawing our attention on Santamore and Cross paper, and A. A. Krushynska for fruitful discussions on dispersion curves in elastic bars of rectangular cross-section. They are grateful for funding provided by LABEX WIFI (Laboratory of Excellence within the French Program Investments for the Future, ANR-10-LABX-24) and ANR-10-IDEX-0001-02 PSL*.\\

\section*{Appendix} 
\appendix
\renewcommand\thesection{A\arabic{section}}
\renewcommand{\thefigure}{A\arabic{figure}} \setcounter{figure}{0}

\section{Displacement field at ZGV resonance frequencies}

\begin{figure*}[ht]
\subfigure{\includegraphics[width=\textwidth]{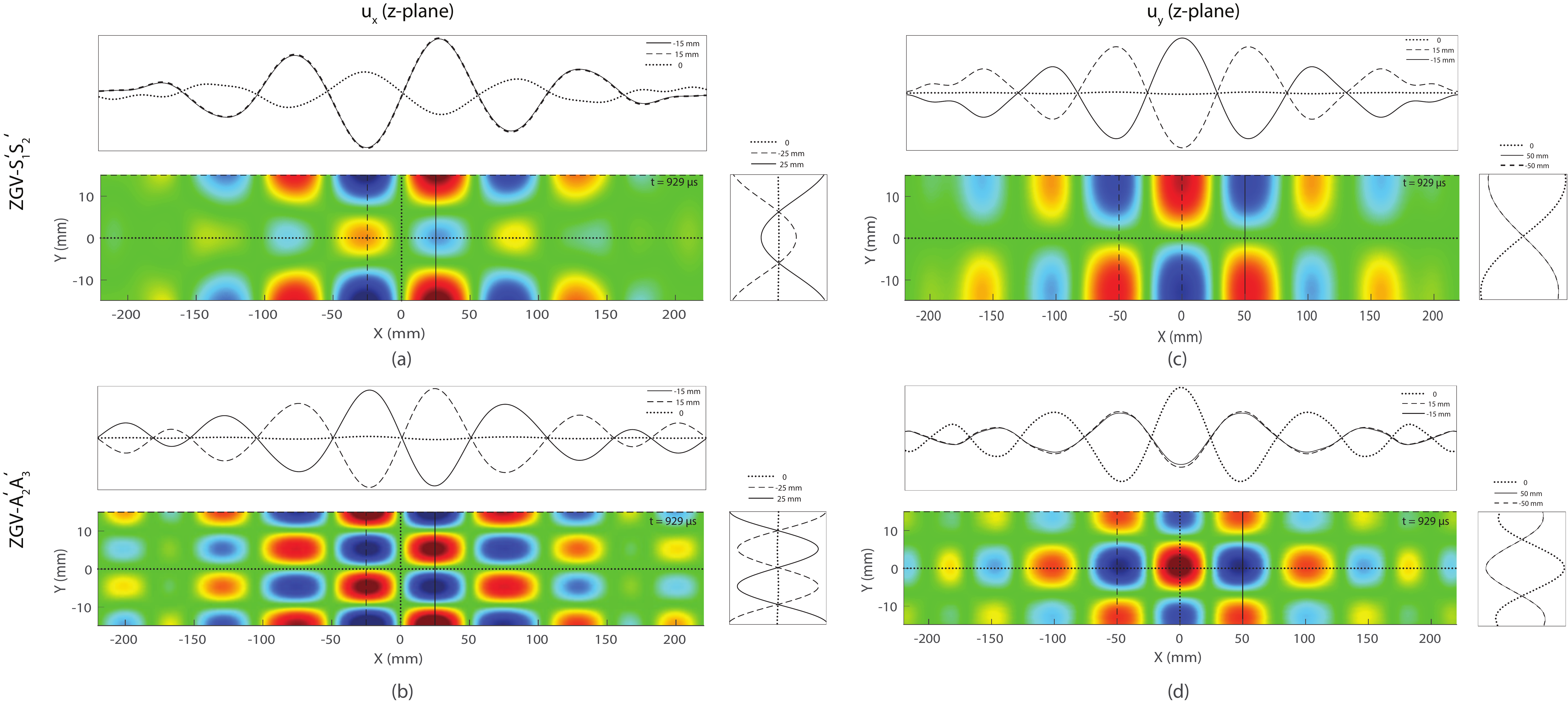}\label{fig:dep5a}}
\subfigure{\label{fig:dep5b}} 
\subfigure{\label{fig:dep5c}}
\subfigure{\label{fig:dep5d}} 
\caption{In-plane displacement components at the $S'_1S'_2$-ZGV frequency $u_x$ (a) and $u_y$ (c) and at the $A'_2A'_3$-ZGV frequency $u_x$ (b) and $u_y$ (d) simulated with the elastodynamic FDTD code Simsonic for a Duraluminum tape of aspect ratio $b/d=15$.}
\label{fig:dep5}
\end{figure*}

For the Duraluminum, the velocity field $(v_x,v_y)$ was also recorded in the plane $z=0$ and the in-plane displacement field deduced by numerical integration. After temporal Fourier transform of $1$~ms signal, the in-plane displacement distribution was observed at the first two ZGV frequencies [Figs.~A\ref{fig:dep5a}-A\ref{fig:dep5d}]. The displacement components $u_x$ and $u_y$ are of the same order of magnitude and similar to those of ZGV Lamb modes.\cite{Balogun07}

\cleardoublepage

\end{document}